\documentclass[aps,prl,letterpaper,superscriptaddress,longbibliography,twocolumn,asmath,amssymb]{revtex4-2}

\usepackage{graphicx}
\usepackage{dcolumn}
\usepackage{bm}
\usepackage{physics}
\usepackage[version=4]{mhchem}
\usepackage{soul}

\usepackage{graphicx}
\usepackage[caption=false]{subfig} 
\usepackage[colorlinks = true,linkcolor = blue,citecolor = blue,urlcolor=blue]{hyperref}
\AtBeginDocument{%
}

\begin{document}

\preprint{APS/123-QED}

\title{$1/3$-Flux Bound States in Multicomponent Superconductor Exhibiting Non-Abelian Statistics of $\mathbb {Z}_3$ Parafermions}

\author{Xing-Yu Wu}
	\affiliation{Interdisciplinary Center for Theoretical Physics and Information Sciences, Fudan University, Shanghai 200433, China}

\author{Y. H. Wang}
    \affiliation{State Key Laboratory of Surface Physics and Department of Physics, Fudan
University, Shanghai 200433, China}
    \affiliation{Shanghai Research Center for Quantum Sciences, Shanghai 201315, China}

\author{X. C. Xie}
    \affiliation{Interdisciplinary Center for Theoretical Physics and Information Sciences, Fudan University, Shanghai 200433, China}
    \affiliation{International Center for Quantum Materials, School of Physics, Peking University, Beijing 100871, China}
    \affiliation{Hefei National Laboratory, Hefei 230088, China}
    
\author{Yijia Wu}
	\thanks{Corresponding author: yijiawu@fudan.edu.cn}
    \affiliation{Interdisciplinary Center for Theoretical Physics and Information Sciences, Fudan University, Shanghai 200433, China}
	\affiliation{State Key Laboratory of Surface Physics, Fudan University, Shanghai 200433, China}
	\affiliation{Hefei National Laboratory, Hefei 230088, China}

\date{\today}

\begin{abstract}
 Multicomponent superconductors (MSCs) are predicted to host magnetic flux quanta carrying arbitrary fractions of the superconducting flux quantum. Recent advances have raised the expectation that $1/3$ flux quanta may emerge in MSCs with $C_3$ rotational symmetry. Electrons bound to such fractional flux are long regarded to form anyons, yet their explicit braiding statistics remain unexplored. We propose that these $1/3$-flux bound states ($1/3$-FBSs) exhibit the intriguing non-Abelian statistics of $\mathbb{Z}_3$ parafermions. Under no-double-occupancy constraint, two successive braiding operations of $1/3$-FBSs are equivalent to a single $\mathbb Z_3$ parafermion braiding operation. The parafermion parity encoding the braiding outcome can be read out via the fermionic occupation number of the $1/3$-FBSs. Combined with a crossed-Andreev-reflection-induced Hadamard gate, one can realize the complete set of $\mathbb{Z}_3$ parafermion braiding operations.
\end{abstract}
\maketitle

\textit{\label{sec:level1}Introduction.}---
The realization of fault-tolerant topological quantum computation (TQC) remains one of the most profound challenges in condensed matter physics and quantum information science~\cite{nayakNonAbelianAnyonsTopological2008}. Thus far, this field has been predominantly driven by the pursuit of Majorana zero modes (MZMs)~\cite{ivanovNonAbelianStatisticsHalfQuantum2001, aliceaNonAbelianStatisticsTopological2011}, whose operators obey the defining relation $\gamma_i^2=1$. As a natural generalization of MZMs, $\mathbb{Z}_d$ parafermions~\cite{greenGeneralizedMethodField1953, zamolodchikovNonlocalParafermionCurrents1985}, whose operators satisfy $\alpha_i^d=1$, host a richer set of non-Abelian braiding statistics. Crucially, they enable the encoding of qudits~\cite{GottesmanFault1999, 10.3389/fphy.2020.589504}, the quantum counterpart of $d$-ary digits, extending the conventional concept of MZM-based qubits to multi-level quantum states. 
Mathematically, the computational utility of MZM-based qubits remains constrained, as their braiding cannot generate the full Clifford group without supplementary resources like measurements~\cite{BondersonMeasurement-Only2008} or ancilla states~\cite{DasSarmaMajorana2015}. In contrast, braiding $\mathbb{Z}_d$ parafermions with odd $d$ intrinsically generates the entire multi-qudit Clifford group~\cite{hutterQuantumComputingParafermions2016}. For prime $d$, this architecture achieves universal quantum computation by adding a single non-Clifford gate~\cite{campbellMagicStateDistillationAll2012, farinholtIdealCharacterizationClifford2014, duaUniversalQuantumComputing2019, benhemouUniversality3Parafermions2023}. As the minimal odd-prime realization, $\mathbb{Z}_3$ parafermions provide a natural route beyond Majorana-based architectures toward richer TQC.

However, conventional proposals for realizing $\mathbb{Z}_d$ parafermions require coupling fractional quantum Hall (FQH) edge states to superconductors~\cite{clarkeExoticNonAbelianAnyons2013, mongUniversalTopologicalQuantum2014, vaeziSuperconductingAnalogueParafermion2014, caoSignaturesParafermionZero2024}, a requirement that presents formidable experimental challenges. Although continuous efforts have been devoted to engineering these hybrid interfaces and inducing robust superconductivity in FQH systems~\cite{JasonopologicalPhaseswithParafermions2016, GulAndreevReflectionintheFractionalQuantumHallState2022}, a definitive experimental demonstration has not yet been achieved. Identifying a more accessible, naturally occurring hardware platform for $\mathbb{Z}_3$ parafermions thus remains a critical open challenge. It is well known that, besides the Moore-Read state in the $\nu=5/2$ FQH system~\cite{MOORE1991}, MZMs also emerge as flux-bound states in $p$-wave superconductors~\cite{ivanovNonAbelianStatisticsHalfQuantum2001}. This observation suggests a broader question: can $\mathbb{Z}_d$ parafermions likewise emerge as bound states of fractional fluxes? Indeed, following Wilczek~\cite{WilczekQuantumMechanicsofFractional-SpinParticles1982}, anyons with fractional statistics have long been understood as composite particles formed by binding electric charges to fractional magnetic fluxes. These considerations motivate the search for $\mathbb{Z}_d$ parafermions in fractional-flux systems.

In this Letter, motivated by recent breakthroughs in the experimental detection of fractional flux in multicomponent superconductors (MSCs)~\cite{iguchiSuperconductingVorticesCarrying2023, zhouObservationSinglequantumVortex2024, zhengObservationQuantumVortex, garciacamposvisualizationvortexsheetshalf2025, GeObservationFluxQuantization2024}, we theoretically demonstrate an experimentally feasible pathway to realizing $\mathbb{Z}_3$ parafermion statistics in MSCs. 
MSCs comprise distinct superconducting components, each characterized by an independent integer winding number, and support fractional vortex excitations beyond conventional topological constraints~\cite{babaevVorticesFractionalFlux2002}. On the experimental front, recent studies establish the realization of these fractional vortices in materials such as \ce{Ba_{1-x}K_xFe2As2}~\cite{iguchiSuperconductingVorticesCarrying2023, zhouObservationSinglequantumVortex2024, zhengObservationQuantumVortex}, \ce{UPt3}~\cite{garciacamposvisualizationvortexsheetshalf2025} and kagome superconductors like \ce{CsV3Sb5}~\cite{GeObservationFluxQuantization2024,cpl_42_3_037401}. The observed temperature-dependent flux, vortex splitting, and fractionalized vortex cores provide compelling evidence for unconventional vortex excitations beyond the conventional Abrikosov paradigm.
The states bound to the fractional vortices observed in the MSCs are  widely regarded as anyon excitations~\cite{TANAKA20091033, zhouObservationSinglequantumVortex2024, timoshukmicroscopic2025}; however, their explicit anyonic braiding properties have not yet been investigated. In this work, we demonstrate that $1/3$-flux bound states ($1/3$-FBSs) in MSCs exhibit the intriguing non-Abelian statistics of $\mathbb{Z}_3$ parafermions. Subject to the no-double-occupancy (NDO) constraint, two successive braiding operations of $1/3$-FBSs are shown to be equivalent to a single $\mathbb{Z}_3$ parafermion braiding operation. Remarkably, we demonstrate that the non-local, experimentally elusive parafermionic parity encoding the braiding outcome can be accessed through measurements of the local and experimentally accessible fermion occupation numbers. Finally, with the assistance of a crossed-Andreev-reflection-induced Hadamard gate, the complete set of $\mathbb{Z}_3$ parafermion braiding operations can be realized.

\textit{$1/3$-Flux Bound States.}---To establish a concrete physical platform, we consider an $n$-component superconductor described by a set of complex order parameters $\Delta_j$'s, where $j=1,2,\ldots,n$. The total supercurrent density is given by~\cite{babaevVorticesFractionalFlux2002} $\bm{J} = \sum_{j=1}^{n} \frac{e}{m_j} n_j (\hbar\nabla\theta_j - 2e\bm{A})$, where $\theta_j$, $n_j$, and $m_j$ denote the phase of the order parameter, the carrier density, and the effective mass of the $j$-th component, respectively. Deep within the superconducting bulk, the supercurrent vanishes $\bm{J} = 0$. Consequently, integrating the magnetic vector potential $\bm{A}$ along a closed loop enclosing the vortex core yields a trapped magnetic flux expressed as a weighted linear combination of the integer phase winding numbers, $w_j = \frac{1}{2\pi}\oint \nabla\theta_j \cdot d\bm{l}$, naturally giving rise to fractionalized flux quanta~\cite{GaraudTopologicalSolitons2011}: $\Phi = \frac{1}{2}\Phi_0 (\sum_{i=1}^n{w_i |\Delta_i|^2)/(\sum_{i=1}^n |\Delta_i|^2}),$
where $\Phi_0 = h/e$ represents the conventional flux quantum, and $\Phi_0/2=h/(2e)$ is the flux quantum in superconducting systems. We consider a symmetric three-component system ($n=3$) with identical pairing amplitudes $|\Delta_1|=|\Delta_2|=|\Delta_3|$. Such conditions are expected to be experimentally accessible in the kagome superconducting systems with $C_3$ rotational symmetry~\cite{GeObservationFluxQuantization2024,JinInterplay2022,Ling-FengZhangKagomeTheory2024}. Under the phase winding configuration $w_1=w_2=1$ and $w_3=0$, the vortex traps a fractionalized magnetic flux:
\begin{equation}
    \Phi = \frac{h}{3e} = \frac{1}{3} \Phi_0.
\end{equation}

Within the cores of these fractional-flux vortices, localized subgap fermionic excitations emerge~\cite{benfenatiboundary2021,benfenatispontaneous2022,timoshukmicroscopic2025}, which we term $1/3$-flux bound states ($1/3$-FBSs). We consider a single $1/3$-flux centered at the origin of a three-component $s$-wave superconductor, described by a Bogoliubov-de Gennes (BdG) Hamiltonian with block-diagonal form $\mathcal{H}(\bm{p}-e\bm{A}_1)= \mathrm{diag}(\mathcal{H}_1, \mathcal{H}_2, \mathcal{H}_3)$. Here, each block $\mathcal{H}_j$ describes the $j$-th superconducting component, where $\mathcal{H}_j(\bm{p}-e\bm{A}_1)$ takes the form of a single-component BdG Hamiltonian with order parameter $\Delta_j$, coupled to the common vector potential $\bm{A}_1=\frac{\Phi}{2\pi r^2}\hat{z}\times\bm{r}$ for the $1/3$-flux $\Phi=h/(3e)$ (see Supplemental Material~\cite{supp} for full solutions of BdG equation). Focusing on the low-energy physics within the superconducting subgap, we isolate the lowest-lying Caroli-de Gennes-Matricon (CdGM) mode near zero energy, denoted as $\Psi_1(\bm{r})$~\cite{caroliBoundFermionStates1964}, and construct the corresponding fermionic creation operator $\psi_1^{\dagger}$, satisfying $\Psi_1(\bm{r})=\bra{\bm{r}}\psi_1^\dagger\ket{0}$.

\begin{figure}[b]
    \centering
    \subfloat[]{\label{fig:1a}\includegraphics[width=0.48\columnwidth]{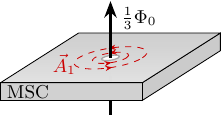}}%
    \hfill
    \subfloat[]{\label{fig:1b}\includegraphics[width=0.48\columnwidth]{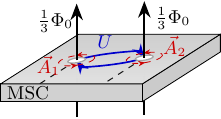}}
    \vspace{0.01cm}
    \subfloat[]{\label{fig:1c}\includegraphics[width=0.68\columnwidth]{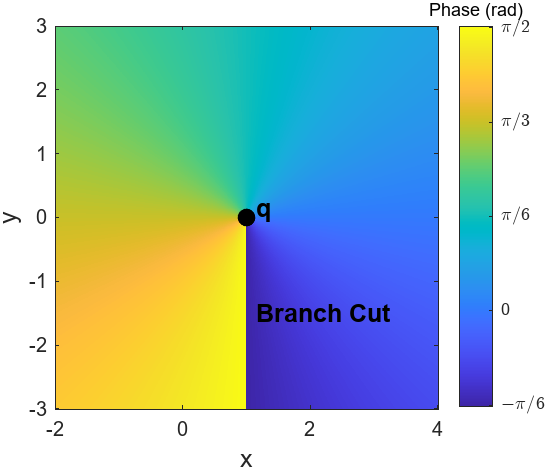}}
    \caption{(a) Schematic of a single fractional vortex carrying magnetic flux $\frac{1}{3}\Phi_0$ in an MSC, accompanied by the circulating vector potential $\bm{A}_1$. (b) Exchange process of two identical 1/3-fractional vortices governed by the braiding operator $U$. Black dashed half-lines extending to the boundary represent the associated branch cuts. (c) Spatial distribution of the gauge field $\Omega(\bm{r})$ associated with the vector potential $\bm{A}_2$ of the second $1/3$-fractional vortex, exhibiting a branch cut along a half-line originating from $\bm{r}=\bm{q}$.}
    \label{fig:vortices}
\end{figure}

The introduction of a second $1/3$-fractional vortex at position $\bm{r}=\bm{q}$ is mathematically equivalent to incorporating an additional gauge field $\bm{A}_2=\frac{\Phi}{2\pi|\bm{r}-\bm{q}|^2}\hat{z}\times(\bm{r}-\bm{q})$, modifying the BdG equation to $\mathcal{H}(\bm{p}-e\bm{A}_1-e\bm{A}_2)\Psi'(\bm{r})=E\Psi'(\bm{r})$. Solving this two-center differential equation directly is analytically intractable due to the breaking of rotational symmetry. However, $\bm{A}_2$ acts as a local pure gauge field near the first vortex for intervortex separations vastly exceeding the superconducting penetration depth. Thus, we map the two-vortex problem onto the single-vortex Hamiltonian via a local $U(1)$ gauge transformation $\mathcal{U} = \exp[i\Omega(\bm{r})(\mathbb{I}_3\otimes\sigma_z)]$, satisfying $\mathcal{U}\mathcal{H}(\bm{p}-e\bm{A}_1-e\bm{A}_2)\mathcal{U}^\dagger = \mathcal{H}(\bm{p}-e\bm{A}_1)$. Here, $\mathbb{I}_3$ is the $3\times3$ identity matrix in the superconducting-component space and $\sigma_z$ acts in the BdG space. This transformation dictates that the two-vortex eigenstates acquire a non-trivial spatial phase factor, related to the original single-vortex wave functions via $\Psi'(\bm{r})=\mathcal{U}^\dagger\Psi(\bm{r})$. To preserve the single-valuedness of the order parameters, the gauge field $\Omega(\bm{r})$ must incorporate a branch cut to properly capture the topological phase winding around the second vortex at $\bm{r}=\bm{q}$, taking the piecewise form:
\begin{equation}
    \Omega(\bm{r}) = \begin{cases}
        \frac{\Phi}{\Phi_0} \arctan\left[\frac{(\bm{r}-\bm{q})\cdot\hat{y}}{(\bm{r}-\bm{q})\cdot\hat{x}}\right], & (\bm{r}-\bm{q})\cdot\hat{x} > 0, \\[6pt]
        \frac{\Phi}{\Phi_0} \left\{ \arctan\left[\frac{(\bm{r}-\bm{q})\cdot\hat{y}}{(\bm{r}-\bm{q})\cdot\hat{x}}\right] + \pi \right\}, & (\bm{r}-\bm{q})\cdot\hat{x} < 0.
    \end{cases}
\end{equation}

In what follows, we focus on the adiabatic exchange of two such $1/3$-FBSs, whose low-energy subspace is governed by the fermionic annihilation operators $\psi_1$ and $\psi_2$. Consequently, as illustrated in Fig.~\ref{fig:1c}, an adiabatic exchange of the two $1/3$-fractional vortices causes one $1/3$-FBS to cross the branch cut of the other, resulting in the accumulation of a nontrivial geometric phase~\cite{ivanovNonAbelianStatisticsHalfQuantum2001, PhysRevLett.125.036801}. The resulting transformation rules for these operators under the unitary braiding operation $U$ read:
\begin{equation}\label{eq:U}
    \begin{cases}
        U\psi_1 U^\dagger = \psi_2, \\
        U\psi_2 U^\dagger = \omega \psi_1.
    \end{cases}
\end{equation}
where $\omega = e^{i 2\pi/3}$. Note that such braiding properties reduce to the Majorana case~\cite{ivanovNonAbelianStatisticsHalfQuantum2001} if $\omega=e^{i\pi}$. To explicitly demonstrate the non-Abelian statistics of these 1/3-FBSs, we express the braiding operator $U$ within the two-fermion Fock space. Choosing the basis $\left\{\ket{00}_\mathrm{f}, \ket{10}_\mathrm{f}, \ket{01}_\mathrm{f},\ket{11}_\mathrm{f}\right\}$, where $\ket{00}_\mathrm{f}$ is the fermionic vacuum state, $\ket{10}_\mathrm{f}=\psi_1^\dagger\ket{00}_\mathrm{f}$, $\ket{01}_\mathrm{f} =\psi_2^\dagger\ket{00}_\mathrm{f}$, and $\ket{11}_\mathrm{f}= \psi_1^\dagger\psi_2^\dagger\ket{00}_\mathrm{f}$, the matrix representation of $U$ reads
\begin{equation}
    U = \begin{pmatrix}
        1 & 0 & 0 & 0 \\
        0 & 0 & 1 & 0 \\
        0 & \omega^2 & 0 & 0 \\
        0 & 0 & 0 & -\omega^2
    \end{pmatrix}.
\end{equation}
Although this operator exhibits non-trivial exchange statistics, it does not directly map onto the standard $\mathbb{Z}_3$ parafermion braiding due to a mismatch in their Hilbert space dimensions. Specifically, while the two-fermion Fock space has a dimension of $2^2=4$, the space spanned by two $\mathbb{Z}_3$ parafermions is $\sqrt{3}^2=3$-dimensional, implying that a physical constraint must be enforced. When strong electronic correlations or Coulomb repulsions are introduced, the double-occupied state $\ket{11}_\mathrm{f}$ incurs a massive energy penalty relative to the remaining three states, effectively projecting it out of the low-energy Hilbert space. To formally incorporate this constraint, we introduce a repulsive interaction $H_{\text{int}} = U' n_1 n_2$ into the effective Hamiltonian, where $n_i=\psi_i^\dagger\psi_i$ ($i=1,2$) is the particle number operator. In the strong-coupling limit ($U' \to \infty$), this term dynamically projects the system into a truncated low-energy Hilbert space by enforcing the no-double-occupancy (NDO) condition $n_1 n_2 = 0$. Evaluating the braiding operator within this projected subspace $\left\{\ket{00}_\mathrm{f}, \ket{10}_\mathrm{f}, \ket{01}_\mathrm{f}\right\}$ yields the effective representation
\begin{equation}\label{eqs:Ue}
    U_{\mathrm{eff}} = \begin{pmatrix}
        1 & 0 & 0 \\
        0 & 0 & 1 \\
        0 & \omega^2 & 0
    \end{pmatrix}.
\end{equation}
This projection remains valid under the adiabatic condition $T \gg \hbar / U'$, where the braiding timescale $T$ is sufficiently long to suppress non-adiabatic excitations into the high-energy double-occupancy state $\ket{11}_\mathrm{f}$.

\textit{Mapping to $\mathbb{Z}_3$ parafermions.} We now explicitly map the braiding statistics of $1/3$-FBSs under the NDO constraint [Eq.~(\ref{eqs:Ue})] to those of $\mathbb{Z}_3$ parafermions. The defining algebra of $\mathbb{Z}_3$ parafermionic modes $\alpha_i$ is given by $\alpha_i^3=1$, $\alpha_i^\dagger=\alpha_i^2$, and $\alpha_i\alpha_j=\omega^{\mathrm{sgn}(j-i)}\alpha_j\alpha_i$, where $\omega = e^{i 2\pi/3}$. The generalized parity operator $\Lambda_{i,i+1}=\alpha_i^\dagger\alpha_{i+1}$ acts on the three-dimensional Hilbert space spanned by two $\mathbb{Z}_3$ parafermions $\alpha_i$ and $\alpha_{i+1}$, defining the parity basis $\left\{\ket{0}_{\mathrm{P}}, \ket{1}_{\mathrm{P}}, \ket{2}_{\mathrm{P}}\right\}$. Here, each eigenstate $\ket{q}_{\mathrm{P}}$, labeled by the $\mathbb{Z}_3$ parity index $q=0,1,2$, satisfies $\Lambda_{i,i+1} \ket{q}_{\mathrm{P}} = \omega^{q+2} \ket{q}_{\mathrm{P}}$. In this basis, the braiding matrix $U_{\mathrm{P}}$ for two neighboring $\mathbb{Z}_3$ parafermions $\alpha_i$ and $\alpha_{i+1}$ takes the diagonal form of (see Supplemental Material~\cite{supp})
\begin{equation}\label{eqs:Up}
    U_{\mathrm{P}} = \begin{pmatrix}
        -i & 0 & 0 \\
        0 & e^{i\frac{\pi}{6}} & 0 \\
        0 & 0 & e^{i\frac{\pi}{6}}
    \end{pmatrix}.
\end{equation}
A direct comparison between Eq.~\eqref{eqs:Ue} and Eq.~\eqref{eqs:Up} establishes an exact algebraic mapping, which constitutes the \emph{central result} of this Letter:
 \begin{equation}\label{eqs:mapping}
    U_\mathrm{P} = -i(U_{\mathrm{eff}}^\dagger)^2.
\end{equation}
This mapping holds under the one-to-one correspondence between the $\mathbb{Z}_3$ parafermion parity basis and the fermionic basis, $\ket{0}_{\mathrm P}\equiv\ket{00}_{\mathrm f}$, $\ket{1}_{\mathrm P}\equiv\ket{10}_{\mathrm f}$, and $\ket{2}_{\mathrm P}\equiv\ket{01}_{\mathrm f}$.
This relation demonstrates that two successive braiding operations of $1/3$-FBSs are equivalent to a single $\mathbb{Z}_3$ parafermion braiding operation (with inverse braiding orientation), up to a global Abelian phase factor of $-i$. The identity $U_{\mathrm{P}}^3=i$ is consistent with the fact that the geometric phase acquired by a $1/3$-FBS is trivial after three traversals of a branch cut, as $\omega^3=1$ [see Eq.~(\ref{eq:U})].
More profoundly, this correspondence implies that non-Abelian $\mathbb{Z}_3$ anyonic statistics do not inherently demand intrinsically fractionalized degrees of freedom. Instead, they can naturally emerge from correlated electronic systems driven entirely by the non-trivial constraints \cite{wangNonAbelianStatisticsBosonic2026}.

This algebraic correspondence allows us to explicitly map the parafermionic operators onto the constrained fermionic degrees of freedom. Using the shift and clock matrices defined as $\sigma\ket{q}_{\mathrm P}=\ket{q+1}_{\mathrm P}$ (mod 3), and $\tau\ket{q}_{\mathrm P}=\omega^q\ket{q}_{\mathrm P}$, we construct a Fradkin-Kadanoff transformation~\cite{FradkinDisorder1980}:
$\alpha_1=\sigma$, and $\alpha_2=\omega^2\sigma\tau$. Under the basis identification $\ket{0}_{\mathrm P}\equiv\ket{00}_{\mathrm f}$, $\ket{1}_{\mathrm P}\equiv\ket{10}_{\mathrm f}$, and $\ket{2}_{\mathrm P}\equiv\ket{01}_{\mathrm f}$ introduced above, $\mathbb Z_3$ parafermions are expressed as
\begin{subequations}\label{eqs:FK}
\begin{align}
    \alpha_1 &= \psi_1^\dagger(1-n_2) + \psi_2^\dagger \psi_1 + (1-n_1)\psi_2, \\
    \alpha_2 &= \omega^2\psi_1^\dagger(1-n_2) + \psi_2^\dagger\psi_1 + \omega(1-n_1)\psi_2.
\end{align}
\end{subequations}
By exploiting the NDO constraint $n_1 n_2 = 0$, the non-local parity operator $\Lambda_{12} = \alpha_1^\dagger \alpha_2$ simplifies to a compact form:
\begin{equation}\label{eqs:parity}
    \Lambda_{12} = \omega^{2+n_1+2n_2} = \omega^{2+n_1-n_2},
\end{equation}
where the second equality naturally follows from the identity $\omega^3 = 1$. Equation~\eqref{eqs:parity} constitutes a further \emph{key result} of this Letter: the nonlocal and experimentally elusive parafermionic parity can be accessed through measurements of the local fermionic occupation operator, namely, $N_{\mathrm{eff}}=n_1+2n_2$ (or equivalently, $N_{\mathrm{eff}}=n_1-n_2$). This correspondence enables the detection of the parafermion parity encoding the braiding outcome via standard solid-state charge-sensing techniques~\cite{SecchiMultidimensional2023, LiuFusion2023, Microsoft_charge_sensing, Kouwenhoven_charge_sensing} measuring local fermion occupation numbers.

\textit{Four $1/3$-FBSs encoding a qutrit.} Because the $\mathbb{Z}_3$ parafermion parity $\Lambda_{i,i+1}$ is conserved under braiding, a minimum of four $\mathbb{Z}_3$ parafermions is required to encode a logical qutrit~\cite{BennettQuantum1999}, the quantum analogue of a ternary digit.  Here, we adopt the sparse-encoding scheme~\cite{ZhanDissipationless2024,FreyMajorana2026}, in which every four $\mathbb{Z}_3$ parafermions encode one logical qutrit. Four $\mathbb{Z}_3$ parafermion operators $\alpha_k$ ($k=1,2,3,4$) are constructed using the standard shift and clock matrices $\sigma$ and $\tau$: $\alpha_1=\sigma\otimes I$, $\alpha_2=\omega^2\sigma\tau\otimes I$, $\alpha_3=\tau\otimes\sigma$, and $\alpha_4=\omega^2\tau\otimes\sigma\tau$, where $I$ denotes the identity operator. Their non-Abelian statistics manifest through the adiabatic exchange of adjacent $\mathbb{Z}_3$ parafermions $\alpha_{i}$ and $\alpha_{i+1}$, described by the elementary braiding operators $U_{\mathrm{P}, i,i+1}=\frac{1}{\sqrt{3}}(1+\alpha_i^\dagger\alpha_{i+1}+\omega\alpha_i^\dagger\alpha_{i+1})$. Within the nine-dimensional Hilbert space spanned by four $\mathbb{Z}_3$ parafermions, all elementary braiding operators $U_{\mathrm{P}, i,i+1}$ are block-diagonal with respect to the total $\mathbb{Z}_3$ parity operator $\Lambda=\Lambda_{12}\Lambda_{34}$, sharing identical matrix representations across all three parity sectors. For instance, in the sector with zero total $\mathbb{Z}_3$ parity index, $Q \equiv q+r \pmod 3 = 0$, spanned by the basis $\left\{\ket{00}_\mathrm{P},\ket{12}_\mathrm{P},\ket{21}_\mathrm{P}\right\}$, where the parity eigenstates $\ket{qr}_\mathrm{P}$ ($q,r \in \{0,1,2\}$) satisfy $\Lambda_{12} \ket{qr}_\mathrm{P} = \omega^{q+2} \ket{qr}_\mathrm{P}$ and $\Lambda_{34} \ket{qr}_\mathrm{P} = \omega^{r+2} \ket{qr}_\mathrm{P}$, the braiding operators read:
\begin{subequations}
\begin{align}
    U_{\mathrm{P},12}= \begin{pmatrix}
            -i & 0 & 0 \\
            0 & e^{i\frac{\pi}{6}} & 0 \\
            0 & 0 & e^{i\frac{\pi}{6}}
        \end{pmatrix},\label{eq:U12}\\
    U_{\mathrm{P},23} = \frac{1}{\sqrt{3}}\begin{pmatrix} 
        1 & \omega^2 & \omega^2 \\ 
        \omega^2 & 1 & \omega^2 \\ 
        \omega^2 & \omega^2 & 1 
        \end{pmatrix},\label{eq:U23}\\
    U_{\mathrm{P},34}=\begin{pmatrix}
            -i & 0 & 0 \\
            0 & e^{i\frac{\pi}{6}} & 0 \\
            0 & 0 & e^{i\frac{\pi}{6}}
        \end{pmatrix},\label{eq:U34}
\end{align} 
\end{subequations}
which explicitly exhibits the non-Abelian nature of $\mathbb{Z}_3$ parafermions, as $U_{\mathrm{P},23}$ does not commute with $U_{\mathrm{P},12}$ and $U_{\mathrm{P},34}$.  

It is natural to utilize four $1/3$-FBSs, labeled by four fermionic annihilation operators $\psi_i$ ($i=1,2,3,4$), to encode a qutrit. Under the NDO constraint $n_1n_2=n_3n_4=0$, the system can be restricted to the logical qutrit subspace spanned by $\left\{\ket{00;00}_\mathrm{f},\ket{10;01}_\mathrm{f},\ket{01;10}_\mathrm{f}\right\}$ in the $Q=0$ parity sector. Here, $\ket{00;00}_\mathrm{f}$ is the fermionic vacuum state, while $\ket{10;01}_\mathrm{f} = \psi_1^\dagger \psi_4^\dagger \ket{00;00}_\mathrm{f}$ and $\ket{01;10}_\mathrm{f} = \psi_2^\dagger \psi_3^\dagger \ket{00;00}_\mathrm{f}$. 
The operators $U_{\mathrm{P},12}$ and $U_{\mathrm{P},34}$ in Eqs.~\eqref{eq:U12} and \eqref{eq:U34} can be implemented by two successive adiabatic exchanges between $1/3$-FBSs $\psi_1$ and $\psi_2$, and between $1/3$-FBSs $\psi_3$ and $\psi_4$, respectively. In contrast, the braiding operator $U_{\mathrm{P},23}$ in Eq.~\eqref{eq:U23} cannot be implemented solely by exchanging $\psi_2$ and $\psi_3$. This limitation arises from two factors: (i) exchanging these two $1/3$-FBSs may violate the underlying NDO constraint in certain sectors, and (ii) a direct adiabatic exchange between them fails to generate the required entanglement among the three logical states.
To realize $U_{\mathrm{P},23}$, an additional operation bringing in entanglement is required. A natural choice for ternary quantum computation is the $\mathbb{Z}_3$ Hadamard gate~\cite{hutterQuantumComputingParafermions2016}
$ H = \frac{1}{\sqrt{3}} \begin{pmatrix}
        1 & 1 & 1 \\
        1 & \omega & \omega^2 \\
        1 & \omega^2 & \omega
    \end{pmatrix}$, 
the $\mathbb{Z}_3$ analogue of the familiar $\mathbb{Z}_2$ Hadamard gate.
By leveraging this algebraic structure, it is straightforward to confirm that the target braiding operator is generated via the similarity transformation
\begin{equation}
    H U_{\mathrm{P},12} H^\dagger = U_{\mathrm{P},23}.
\end{equation}
This identity demonstrates that the braiding operation $U_{\mathrm{P},23}$ in the four-parafermion system can be implemented with the assistance of the $\mathbb{Z}_3$ Hadamard gate. Moreover, because the relations above are valid in all three parity sectors, the resulting gate implementation is identical across different parity sectors.

\textit{Discussion.}---In the system of four $1/3$-FBSs, the requisite $\mathbb{Z}_3$ Hadamard gate can be generated by an effective Hamiltonian $h_{\mathrm{eff}}$ satisfying $e^{-ih_{\mathrm{eff}} \Delta t}=H$, which takes the form
\begin{equation}\label{eq:Heff}
\begin{aligned}
h_{\text{eff}} = E_0 &+ \sum_{i=1}^4 \varepsilon \psi_i^\dagger \psi_i + \Delta\left(\psi_1^\dagger \psi_4^\dagger + \psi_2^\dagger \psi_3^\dagger + \text{H.c.}\right) \\
&+ J\left(\psi_1^\dagger \psi_4^\dagger \psi_2 \psi_3 + \text{H.c.}\right).
\end{aligned}
\end{equation}
The effective Hamiltonian $h_{\mathrm{eff}}$ [Eq.~\eqref{eq:Heff}] describes the low-energy physics of two-dimensional $s$-wave vortex arrays. The bound-state energies $\varepsilon$ of the $1/3$-FBSs are controlled by local chemical potentials, while the non-local pairing amplitude $\Delta$ describes crossed Andreev reflection (CAR) processes between $1/3$-FBSs $\psi_1$ and $\psi_4$, and between $\psi_2$ and $\psi_3$~\cite{Hofstetter2009, zatelli2026majorana}. The four-fermion interaction term $J$ originates from higher-order cotunneling processes. The required parameter ratios $\varepsilon/\Delta=-3/4$ and $J/\Delta=-1/2$~\cite{supp} are experimentally accessible by tuning the intervortex spacing relative to the superconducting coherence length. Evolution under $h_{\mathrm{eff}}$ for a duration $\Delta t=\sqrt{3}\pi/(8\varepsilon)$ then implements the $\mathbb{Z}_3$ Hadamard gate.

By leveraging the $\mathbb{Z}_3$ Hadamard gate, our scheme generates the complete set of $\mathbb{Z}_3$ braiding operations while circumventing the experimental challenges associated with conventional $\mathbb{Z}_3$ parafermion constructions based on FQH edge states coupled to superconductors. 
Combined with direct readout through local fermion occupation measurements, this approach provides a flexible framework for implementing $\mathbb{Z}_3$ parafermion topological quantum gates and directly accessing their braiding outcomes. Furthermore, Eq.~\eqref{eqs:mapping} implies $U_{\mathrm{eff}}^\dagger=(iU_{\mathrm P})^{1/2}$, revealing that a single exchange of two $1/3$-FBSs implements an operation corresponding to a half braid of $\mathbb{Z}_3$ parafermions, beyond the elementary braiding operations conventionally available in $\mathbb{Z}_3$ parafermion systems.

Finally, generalizing this framework to $\Phi_0/d$ vortices hosting $\mathbb{Z}_d$ parafermions encounters a fundamental dimensional constraint. Two localized fermionic modes span a $2^2 = 4$-dimensional Hilbert space, which is reduced to three dimensions under the NDO constraint, exactly matching the Hilbert space of a $\mathbb{Z}_3$ parafermion pair. For $d\geq5$, two fermionic modes cannot accommodate the required $d$-dimensional Hilbert space ($2^2<d$), necessitating additional vortices or internal degrees of freedom. 
Thus, the $1/3$-FBS construction is a special realization of $\mathbb{Z}_3$ topological quantum computation that does not straightforwardly extend to arbitrary $\mathbb{Z}_d$ parafermions.

\textit{Acknowledgement}---
We thank Hong-Yu Wang for fruitful discussion. This work is financially supported by the National Key R\&D Program of China (Grant No. 2024YFA1409000), the Quantum Science and Technology-National Science and Technology Major Project (Grant No. 2021ZD0302400), the National Natural Science Foundation of China (Grant No. 12304194, and No. 12574171), Shanghai Municipal Science and Technology (Grant No. 24DP2600100), Shanghai Pilot Program for Basic Research - Fudan University 21TQ1400100 (25TQ003), and Shanghai Science and Technology Innovation Action Plan (Grant No. 24LZ1400800).


\bibliography{refs}

\end{document}